\RequirePackage[2020-02-02]{latexrelease}
\documentclass[aps,twocolumn,floats,nofootinbib]{revtex4} 

\usepackage{graphics,graphicx,epsfig, sidecap}
\usepackage{graphicx} 
\usepackage{sidecap}
\usepackage{amssymb,color} 
\usepackage{epsf,epstopdf,wrapfig} 
\usepackage {amsmath} 
\usepackage{textgreek} 
\usepackage{sidecap} 
\usepackage{jabbrv} 
\newcommand{\beq}{\begin{equation}} 
\newcommand{\eeq}{\end{equation}} 
\newcommand{\beqn}{\begin{eqnarray}} 
\newcommand{\eeqn}{\end{eqnarray}}

\begin{document} 
	
	\title{Successes and failures of simple statistical physics models for a network of real neurons}

	\author{Leenoy Meshulam$^{1,2}$, Jeffrey L. Gauthier$^{8}$, Carlos D. Brody$^{3,6,7}$, David W. Tank$^{3,4,6}$, and William Bialek$^{4,5,9}$} 
	
	\affiliation{$^1$Center for Computational Neuroscience, and $^2$Department of Applied Mathematics, University of Washington, Seattle, Washington 98195}
	\affiliation{$^3$Princeton Neuroscience Institute,  $^4$Joseph Henry Laboratories of Physics, $^5$Lewis--Sigler Institute for Integrative Genomics, $^6$Department of Molecular Biology, and $^7$Howard Hughes Medical Institute, Princeton University, Princeton, NJ 08544}
	\affiliation{$^8$Department of Biology, Swarthmore College, Swarthmore, Pennsylvania 19081}
	\affiliation{$^9$Initiative for the Theoretical Sciences, The Graduate Center, City University of New York, 365 Fifth Ave., New York, NY 10016}

\begin{abstract}
Biological networks exhibit  complex,  coordinated patterns of activity.  Can these patterns be captured precisely  in simple models?  Here we use measurements of simultaneous activity in 1000+ neurons in the mouse brain to test the validity of  models grounded in statistical physics. When cells are dense samples from a small region, we find extremely detailed quantitative agreement between theory and experiment; sparse samples from larger regions lead to model failures. These results show we can aspire to more than qualitative agreement between simplifying theoretical ideas and the detailed behavior of a complex biological system.
\end{abstract}

\date{\today}

\maketitle

\textit{Introduction.}
In statistical physics we routinely capture the behavior of complex systems, quantitatively,  with models that are much simpler than the underlying microscopic mechanisms.  Can we expect the same level of success in describing biological systems?  Maximum entropy methods \cite{jaynes1957information, jaynes1982rationale} provide a direct path from data to simplified statistical physics models, and this has been used in  systems ranging from protein families to flocks  of birds \cite{schneidman2006weak,tkacik2006ising, tkacik2009spin, ohiorhenuan2010sparse, mora2010maximum, marks2011protein, lapedes2012using, bialek2012statistical, granot2013stimulus, bialek2014social, asti2016maximum, bitbol2016inferring, tavoni2017functional, russ2020evolution}. In particular, these models accurately describe the activity distribution across populations of $N \sim 100$ cells in the dorsal hippocampus of mice \cite{meshulam2017collective}.  But it is unclear whether these successes are significant: some authors claim these models  describe only weak correlations \cite{roudi2009pairwise,macke2012biased}, while others worry that they are complex enough to  make  their success uninformative.

Recent progress in experimental methods creates new opportunities to test these theoretical ideas.   Experiments now monitor 1000+ neurons in the hippocampus, which means we can choose groups of $N=100$ cells in many different ways. We find that  maximum entropy models which match pairwise correlations accurately predict higher order structure in the activity of spatially contiguous local subgroups, but not in distant subgroups where we draw the cells at random from a larger area. This aligns with evidence for spatial organization of neural activity in dorsal hippocampus \cite{wiener1989spatial, hampson1999distribution, rickgauer2014simultaneous}. Importantly, in the most successful examples all of the higher order statistical features of network activity that we test are predicted within experimental error. This sets a high standard for what we mean when we say that a model ``works.''

\begin{figure*}[!ht]
	\centering
	\includegraphics{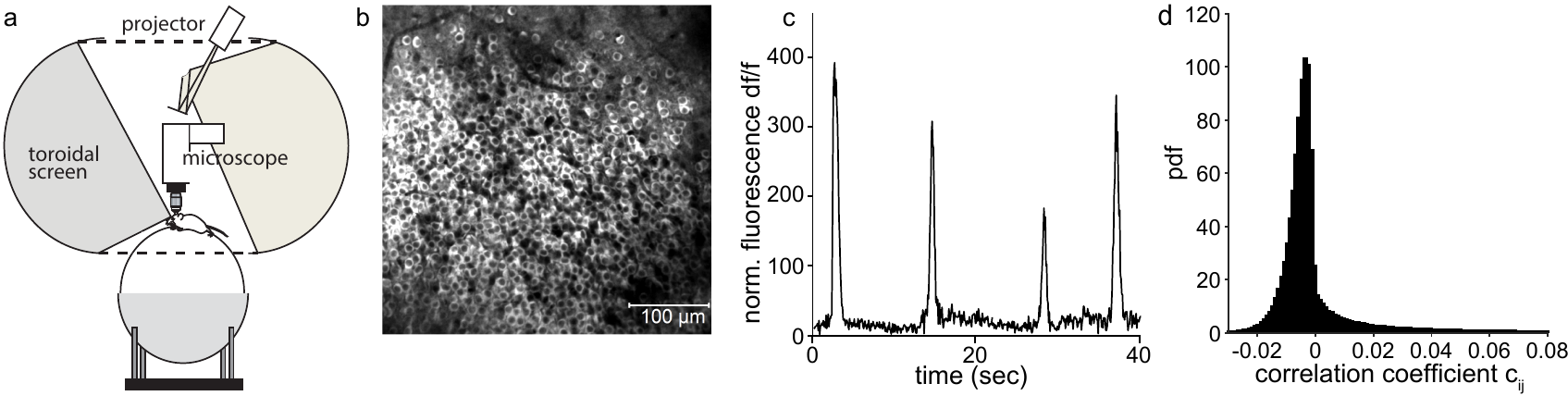}
	\caption{\textbf{Experimental background.} \textbf{(a)} Schematic.  A two--photon microscope is used to image large  neuronal populations in a head-fixed mouse. Feedback from the animal running on the ball advances the virtual corridor projected on the toroidal screen. \textbf{(b)} Fluorescence image of 1485 neurons expressing calcium-sensitive fluorescent protein (averaged). 
	\textbf{(c)} Sample trace of fluorescence  from a single CA1 neuron, showing the high SNR. Raw data were motion corrected, normalized, and binarized \cite{meshulam2017collective}. \textbf{(d)} Probability distribution of correlation coefficients, $c_{\rm ij} = {C_{\rm ij}}/{\sqrt{C_{\rm ii} C_{\rm jj}}}$,  across all pairs of neurons in  \textbf{(b)}.}
	\label{setup}
\end{figure*}

\textit{Strategy.}
Experiments analyzed here monitor the activity of many neurons in the CA1 region of the mouse hippocampus  \cite{gauthier2018dedicated,gauthier2018SEUDO}. Briefly, mice have been genetically engineered to express GCaMP3, a calcium--sensitive fluorescent protein, and the resulting fluorescence is measured using a scanning two--photon microscope.  The mouse runs on a floating ball while its head is fixed, and rotation of ball is fed back to a visual stimulus to create the virtual experience of running along a $4\,{\rm m}$ track (Fig 1a).  We focus on a data set with  $N_{\rm total}=1485$ neurons (Fig \ref{setup}b).  Each experiment generates a $T \sim 40\,{\rm min}$ time series of the fluorescence signal from each neuron (Fig~\ref{setup}c), with frames of duration $\Delta \tau = 1/30\,{\rm s}$. After denoising  we binarize each neuron's activity in each frame, based on whether it was active or silent, $\sigma_{\rm i} \equiv \{0,1\}$  \cite{meshulam2017collective}.

Neurons in the hippocampus include  ``place cells'' that are active only when the animal visits a particular location \cite{okeefe1978hippocampus, okeefe1979review, dombeck2010functional}.  The network of place cells  is thought to form a ``cognitive map,''  contributing to the animal's navigation ability. In rodent  CA1,   $30-50\%$ of neurons are  place cells in any single experimental environment \cite{dombeck2010functional, gauthier2018dedicated, meshulam2017collective}.

\begin{figure} 
		\centering 
		\includegraphics{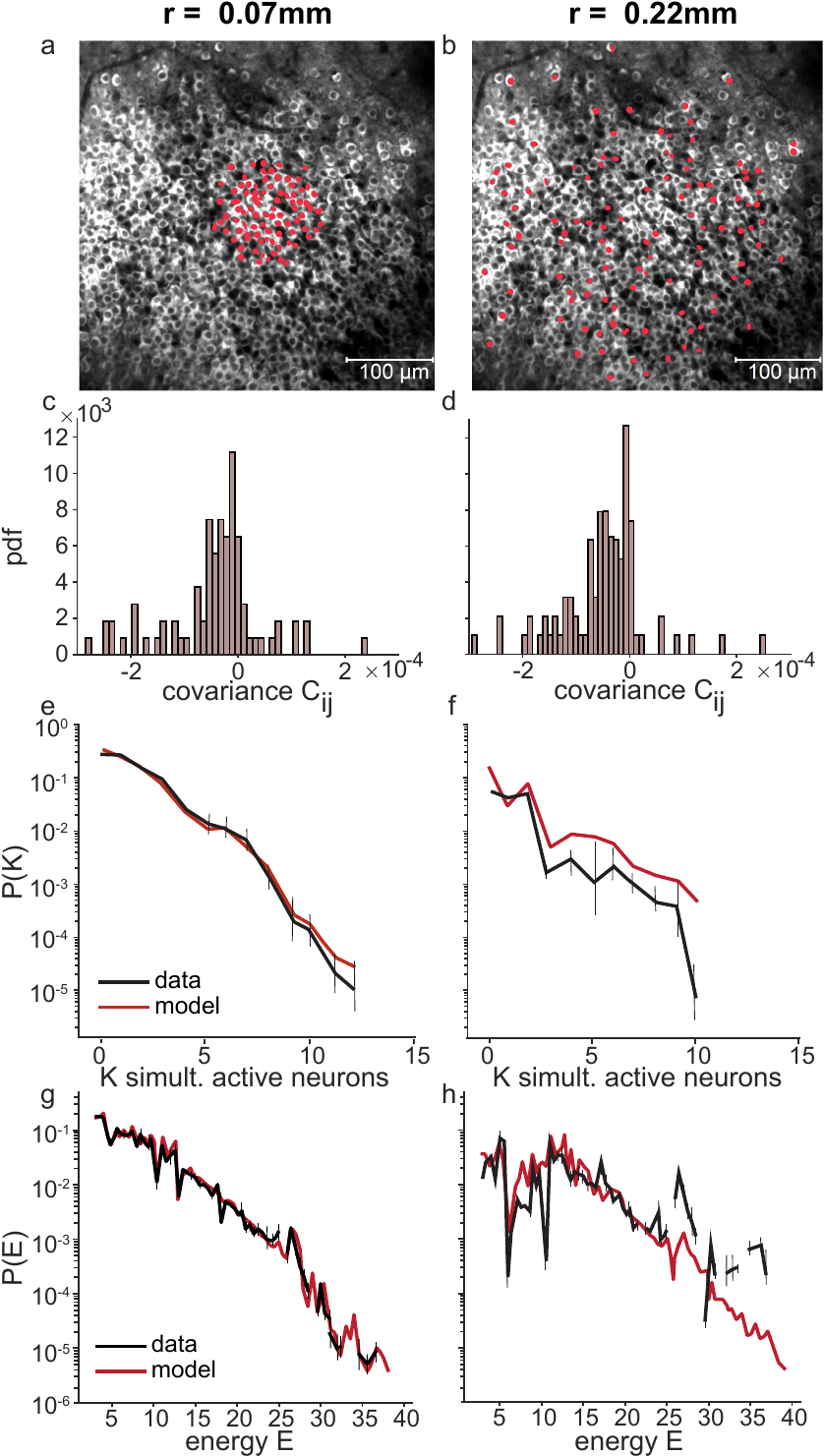}
	    \caption{\textbf{Theory and experiment for ``local'' (left: a, c, e, g) and ``distant'' (right: b, d, f, h) subgroups}. \textbf{(a, b)} Fluorescence image of the entire field of view, as in Fig \ref{setup}b. Red marks are the cells included in example local \textbf{(a)} and distant  \textbf{(b)} subgroups. \textbf{(c, d)} Probability distributions of covariances $C_{\rm ij}$ for the  subgroups  in {\bf (a)} and  {\bf (b)}, respectively. \textbf{(e, f)} The probability that $K$ out of the $N=100$ neurons in the subgroup population are active simultaneously.   \textbf{(g, h)} The distribution of effective energies, $P(E)$, across all network states. In \textbf{(e--h)},  predictions in red, mean and  standard deviations over random halves of the data in black.
	    \label{predictions1}}
\end{figure}

\textit{Assembling subgroups.} To begin we  choose a cell at random and draw a circle of radius $r=0.07\,{\rm mm}$.  This circle contains $N \sim 100$ cells,  simulating  previous generation of experiments with a more limited field of view \cite{meshulam2017collective}, and  we refer to these cells  as a ``local'' subgroup of neurons.   With the same cell at the center we draw larger circles---$r \sim  0.11,\, 0.14,\, 0.18,\, 0.22\,{\rm mm}$, increasing $2\times$ in area at each step---and then choose cells at random from these regions.  For different groups to be comparable we want to keep the fraction of place cells fixed; this keeps the distribution of mean activities $\langle\sigma_{\rm i}\rangle$ across the group roughly constant as well.  Because the models we study  are built by matching pairwise covariances, 
\begin{equation}
    C_{\rm ij} = \langle \left(\sigma_{\rm i} - \langle\sigma_{\rm i}\rangle\right) \left(\sigma_{\rm j} - \langle\sigma_{\rm j}\rangle\right) \rangle ,
\label{Cij_def}
\end{equation}
we would also like to fix the distribution of these correlations. At radius $r_k$ the distribution of  matrix elements is $P_k(C)$; to expand $r_k \rightarrow r_{k+1}$ we swap cells in the smaller radius for those in the larger radius, and at every swap we minimize the Kullback--Leibler divergence between $P_k(C)$ and $P_{k+1}(C)$, until half of the cells have been swapped.  This generates a nested collection of groups  from progressively larger regions of the same brain area, matched in their basic statistical properties; Fig \ref{predictions1}a shows an example of the smallest (``local'') and largest (``distant'') groups.  These subgroups also simulate experiments that randomly sample a fraction of the cells in a region, down to $\sim 10\%$ at $r = 0.22\,{\rm mm}$. 

\textit{Maximum entropy models for subgroups.}
In each of the subgroups we are faced with a population of $N = 100$ neurons, and in each small window of time ($\Delta\tau = 33\, {\rm ms}$) every cell either is active ($\sigma_{\rm i} =1$) or silent ($\sigma_{\rm i} = 0$). The state of the entire network then is an $N$--bit binary vector $\{\sigma_{\rm i}\} = \{\sigma_1,\, \sigma_2,\, \cdots ,\, \sigma_N\}$.  A theory of the network should  predict the probability that we find the system in any one of these states, $P(\{\sigma_{\rm i}\})$, telling us how surprised we should be by any pattern of activity  in the network.

The maximum entropy method  constructs approximations to the distribution  $P(\{\sigma_{\rm i}\})$ that have as little structure as possible, or equivalently generate network states that are as random as possible, while matching some observed properties of the system   \cite{jaynes1957information, jaynes1982rationale}.  It is a theorem that the only way to make ``as random as possible''  mathematically precise is to maximize the entropy \cite{shannon1948mathematical},
\begin{equation}
S = - \sum_{\{\sigma_{\rm i}\}} P(\{\sigma_{\rm i}\}) \log  \left[ P(\{\sigma_{\rm i}\}) \right].
\label{entropy_def}
\end{equation}
There is no unique choice for which properties of the system we should match.   For a network of neurons, it makes sense to match the mean activity of every neuron, 
\begin{equation}
    \langle\sigma_{\rm j}\rangle \equiv \sum_{\{\sigma_{\rm i}\}} P(\{\sigma_{\rm i}\}) \sigma_{\rm j} = \langle\sigma_{\rm j}\rangle_{\rm expt} ,
    \label{match1}
\end{equation}
where $\langle\cdots\rangle_{\rm expt}$ is the average across the experiment.  As a first step toward capturing the interactions among neurons in the network, we  match the correlations between the activity in all pairs of cells $\rm jk$, 
\begin{equation}
    \langle\sigma_{\rm j}\sigma_{\rm k}\rangle \equiv \sum_{\{\sigma_{\rm i}\}} P(\{\sigma_{\rm i}\}) \sigma_{\rm j}\sigma_{\rm k} = 
    C_{\rm jk}^{\rm expt} + \langle\sigma_{\rm j}\rangle_{\rm expt}\langle\sigma_{\rm k}\rangle_{\rm expt} .
    \label{match2}
\end{equation}

The probability distribution that maximizes the entropy in Eq (\ref{entropy_def}) while obeying the constraints in Eqs (\ref{match1}) and (\ref{match2}) has the form
\begin{eqnarray}
 P(\{\sigma_{\rm i}\}) &=& \frac{1}{Z}\exp[- E(\{\sigma_{\rm i}\})]
  \label{model1}\\
E(\{\sigma_{\rm i}\}) &=& -\sum_{{\rm i} =1}^{N} h_{\rm i}\sigma_{\rm i} -\frac{1}{2}\sum_{{\rm i},{\rm j}=1}^{N}J_{\rm ij}\sigma_{\rm i}\sigma_{\rm j }.  
\label{model2}\\
Z &=& \sum_{\{\sigma_{\rm i}\}} \exp[- E(\{\sigma_{\rm i}\})].
\label{model3}
 \end{eqnarray}
This is  the statistical mechanics  of Ising spins in magnetic fields $h_{\rm i}$ with couplings $J_{\rm ij}$ \cite{mezard1988spin}.  To complete the construction we have to find the $\{h_{\rm i}, J_{\rm ij}\}$.  We use Monte Carlo  to draw samples out of the distribution in Eqs (\ref{model1}--\ref{model3}), compute the expectation values $\langle\sigma_{\rm j}\rangle$ and $\langle\sigma_{\rm j}\sigma_{\rm k}\rangle$, and adjust  $\{h_{\rm i}, J_{\rm ij}\}$ until Eqs (\ref{match1}) and (\ref{match2}) are satisfied \cite{meshulam2017collective}.

A crucial feature of the maximum entropy method is that once we have matched the  measured mean activities and pairwise correlations, there are no free parameters. Thus we make {\em parameter--free} predictions for all the higher order statistical structure in the network.  
We have examples where these predictions have been successful  \cite{granot2013stimulus, schneidman2006weak,meshulam2017collective}, but this success is not guaranteed \cite{ganmor2011sparse, tkavcik2014searching}.

\textit{Scope.}  
For each dataset we have chosen ten different neurons as the center for  the construction of subgroups described above, then built maximum entropy models for  local  and distant subgroups of $N=100$ cells.   We focus on  one typical example; populations at intermediate radii interpolate between the extremes described here, and   results are consistent across all examples from three different mice.  We look  at six different predictions, contrasting the levels of success in local and distant subgroups.

\textit{Simultaneous activity.} 
The maximum entropy model aims to capture the {\em collective} nature of population activity. A first test  is to ask for the probability that $K$ out of the $N$ neurons are active in the same small window of time, $P(K)$. Figure \ref{predictions1}e shows this distribution for the local subgroup, and  Fig \ref{predictions1}f shows the same distribution for the distant one.  The local subgroup predictions stay within experimental error bars throughout the entire range, even down to probabilities $10^{-4}-10^{-5}$ at $K>10$. In contrast, the distant subgroup predictions are out of measurement error bounds for all  $K>3$.

\textit{Density of states.}  The fundamental object of our models is the effective energy $E(\{\sigma_{\rm i}\})$ assigned to every network state. 
We can look at the distribution $P(E)$  across the experimentally observed states, as well as the distribution predicted from the theory itself.   Predictions are in excellent quantitative agreement with experiment  for the local subgroup (Fig ~\ref{predictions1}g), including details that one might be tempted to dismiss as noise but in fact are significant.  Agreement extends to $E>30$, corresponding to individual network states with relative probabilities of $\sim 10^{-13}$.   In contrast, predictions in the distant subgroup fail already at $E\sim 7$ (Fig ~\ref{predictions1}h).

\begin{figure} 
	\includegraphics{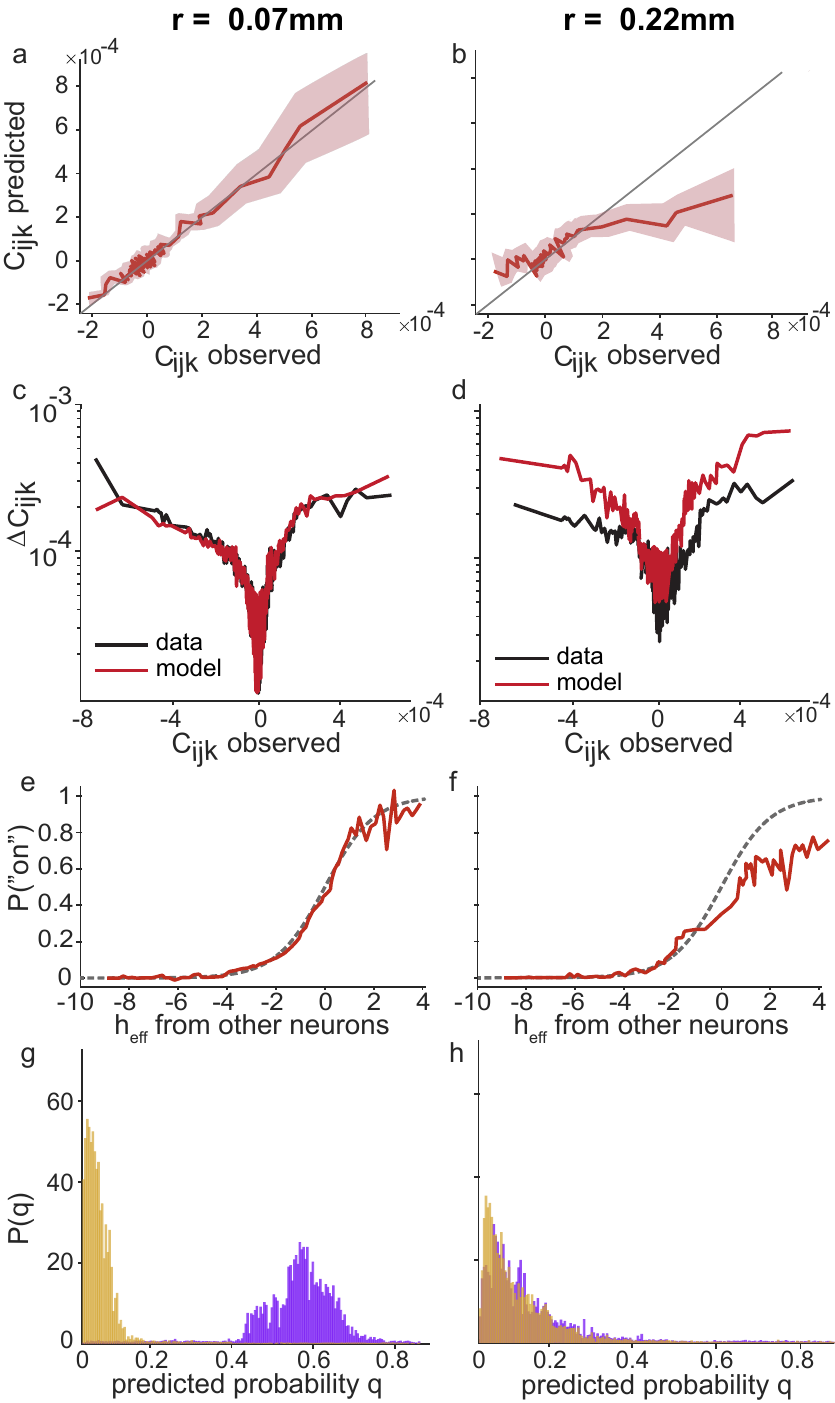} 
	\caption{\textbf{Triplets and effective fields in local  (left: a, c, e, g) and distant (right: b, d, f, h) subgroups.}  \textbf{(a, b)} Comparison of the maximum entropy model predictions for  triplet correlations, $C_{\rm ijk}$, with experiment. All $N^3 / 3\! \sim 10^5$ values are binned, adaptively, along the x--axis, showing mean (red line) and standard deviation (shading). \textbf{(c, d)} Root--mean--square errors of individual predictions (red) and measurements (black). \textbf{(e, f)} Probability of neuron to be active $\sigma_{\rm i} = 1$ vs effective field $h_{\rm i}^{\rm eff}$.
Data in red, parameter--free prediction from Eq (\ref{heff_def}) shown as dashed black line. \textbf{(g, h)} Conditional probability of effective field, expressed as $q$ from Eq (\ref{logit}), given the state of the neuron, $P(h_{\rm i}^{\rm eff}| \sigma_{\rm i})$, with $\sigma_{\rm i} =1$ in purple and $\sigma_{\rm i} =0$ in yellow.
	\label{predictions2}}
\end{figure}

\textit{Triplet correlations.}  
We have constructed models that match correlations between pairs of  neurons, so  a natural test of the theory is to  predict correlations among  triplets,
\begin{equation}
C_{\rm ijk}\equiv \langle (\sigma_{\rm i} -\langle\sigma_{\rm i}\rangle) (\sigma_{\rm j}-\langle\sigma_{\rm j}\rangle)(\sigma_{\rm k}-\langle\sigma_{\rm k}\rangle)\rangle .
\label{C3}
\end{equation}
For $N=100$ cells there are $\sim 1.6 \times 10^5$ distinct triplets. In Figs \ref{predictions2}a and b we group these into bins, showing the mean and standard deviation in each bin, plotting measured vs predicted values. For the local subgroup, predictions are within experimental error across the full dynamic range of the data (Fig \ref{predictions2}a).  For the distant subgroup, in contrast, there is  a dramatic mismatch between theory and experiment  (Fig \ref{predictions2}b).

\textit{Individual triplets.} As a more stringent test we scrutinize each of the  $\sim 1.6 \times 10^5$  triplets individually.  Concretely, we compute the root-mean-square (rms) difference between each predicted and measured correlation (inside small bins) as well as the rms error in our measurements. For the local subgroup, prediction errors track measurement errors  (Fig \ref{predictions2}c); it is almost impossible to do better without over--fitting. Errors in the distant subgroup are much larger, including for the smallest and most frequent values of  $C_{\rm ijk}$ (Fig \ref{predictions2}d). 

\textit{Effective fields.} As usual with Ising models we can define an effective field $h_{\rm i}^{\rm eff}$, 
\begin{eqnarray}
h^{\rm eff}_{\rm i} &=& E ( \sigma_1, \, \cdots ,\,\sigma_{\rm i}=0, \,\cdots , \,{\sigma_N} ) 
\nonumber \\
&& 
\,\,\,\,\,\,\,\,\,\, 
- E ( \sigma_1, \, \cdots ,\,\sigma_{\rm i}=1, \,\cdots , \,{\sigma_N} )\\
&=& h_{\rm i} + \sum_{{\rm j}\neq{\rm i}} J_{\rm ij}\sigma_{\rm j} ,
\label{heff_def}
\end{eqnarray}
which predicts how the network state influences probability that a single neuron is active,
\begin{equation}
q \equiv P(\sigma_{\rm i} =1| h^{\rm eff}_{\rm i} ) = \frac{1}{1+\exp(-h^{\rm eff}_{\rm i})}.
\label{logit}
\end{equation}
In Figs \ref{predictions2}e and f we test these predictions: we obtain an effective field value for each neuron at every moment in time given the state of the rest of the population, and then pool all of these values together. Each  panel depicts the probability of a neuron to be active against the effective field of the same neuron, on top of the prediction from Eq (\ref{logit}). Again the agreement between theory and observations is very good for the local subgroup (Fig \ref{predictions2}e), even for very large effective fields, while prediction quality drops in the distant subgroup (Fig \ref{predictions2}f).

\textit{Inferring the field.} Does a neuron being active (silent) predict a high (low) value of the effective field? This is related to Eq (\ref{logit}) via Bayes rule:
\begin{equation}
P(h_{\rm i}^{\rm eff} | \sigma_{\rm i}) 
= {{P(\sigma_{\rm i}  | h_{\rm i}^{\rm eff}  )P(h_{\rm i}^{\rm eff})}\over{P(\sigma_{\rm i})}}.
\end{equation}
There is no ``correct'' distribution of $h_{\rm eff}$, but it would be attractive if neurons being active predicted that the whole network is in a state that favors this activity [large $q$ in Eq (\ref{logit})]; conversely silent neurons should predict small  $q$.  Figures \ref{predictions2}g and h show the distributions $P(q|\sigma)$, assembled by sampling all  moments in time when a neuron was active or silent.  A clear separation between the two distributions indicates that our model really does distinguish network states that drive activity or silence of individual cells.   This works for the local subgroup (Fig \ref{predictions2}g) but not for the distant subgroup (Fig \ref{predictions2}h).

\textit{Discussion.}   It has never been clear whether relatively simple statistical physics models for networks of neurons should be taken seriously as quantitative theories.  The class of models we have considered is based on the  idea that all structure in neural activity is encoded in the matrix of pairwise correlations; all subsequent predictions (as in Figs \ref{predictions1} and \ref{predictions2}) are parameter--free.  We draw three main conclusions.  First, parameter--free predictions of simple theories can agree with real biological data in surprising quantitative detail: for local subgroups, all predictions we test agree with experiment within measurement errors.   Second, this  is not guaranteed:  we can choose different groups of $N=100$ neurons from the same area of the brain, with similar patterns of mean activity and pairwise correlations in response to the same behaviors, and predictions of the simple model fail.   These results are reproduced in thirty different local/distant subgroup pairs across three different mice.

A corollary  is that we need to be careful in saying that we have agreement or disagreement between theory and experiment. Looking only at distant subgroups  one might be tempted to say that we capture trends in the data, and that the lack of quantitative agreement is the result of over--simplification.  But this is not the case, since we achieve detailed quantitative agreement using the same theory to describe a different group of cells in the same area under the same conditions. This leads to a third conclusion, that success or failure of our predictions reflects properties of the underlying network.  Success is found when we describe a densely sampled collection of cells from a contiguous region,  suggesting that effective interactions are  local; drawing a tight circle around $N=100$ cells we come closer to capturing a complete network.

Work supported in part by the National Science Foundation through the Center for the Physics of Biological Function (PHY--1734030); by the Simons Collaboration on the Global Brain;  by the Howard Hughes Medical Institute; and by the Swartz Foundation.
This paper is dedicated to the memory of Cristina Domnisoru, a luminary who inspired so many; a continuing source of brilliance, creativity, and wonder. Cristina burned selflessly and incisively to better the world, and leaves a lasting echo in her wake.

\bibliography{simple-models2}

\end{document}